\documentclass[12pt,reqno]{article}
\usepackage{amsfonts}
\usepackage{amsmath}
\usepackage{amsbsy}

\usepackage{amssymb,latexsym}
\numberwithin{equation}{section}

\begin{document}
 \allowdisplaybreaks[1]
\title{Integrated Lax Formalism for PCM}
\author{Nejat T. Y$\i$lmaz\\
Department of Electrical and Electronics Engineering,\\
Ya\c{s}ar University,\\
Sel\c{c}uk Ya\c{s}ar Kamp\"{u}s\"{u},\\
\"{U}niversite Caddesi, No:35-37,\\
A\u{g}a\c{c}l{\i} Yol, 35100,\\
Bornova, \.{I}zmir, Turkey.\\
\texttt{nejat.yilmaz@yasar.edu.tr}} \maketitle
\begin{abstract}
By solving the first-order algebraic field equations which arise
in the dual formulation of the $D=2$ principal chiral model (PCM)
we construct an integrated Lax formalism built explicitly on the
dual fields of the model rather than the currents. The Lagrangian
of the dual scalar field theory is also constructed. Furthermore
we present the first-order PDE system for an exponential
parametrization of the solutions and discuss the Frobenious
integrability of this system.
\end{abstract}

\section{Introduction}
As an integrable system the $D=2$ principal chiral model (PCM)
\cite{book,mik} has been extensively studied both in physics and
mathematics literature. PCM in general is an example of a sigma
model \cite{gellmann} with a lie group target space (G). From the
integrability point of view it can be formulated as the
compatibility or the integration conditions (flat curvature
conditions for the relative current) of a pair of linear matrix
equations and some additional constraint which fixes the gauge
among the other integrable systems which can also be derived
within the same context \cite{mik}. For this reason it has gauge
relations to the other integrable systems. Alternatively it can
directly be derived from again the zero curvature integrability
condition of a pair of linear equations known as the Lax pair via
an introduction of a complex spectral parameter. From geometrical
point of view it corresponds to the theory of (pseudo)-harmonic
maps from (pseudo)-Riemannian manifolds into compact Lie groups
\cite{uhlen}. Certain class of fermionic models can also be
formulated as PCM \cite{zakmik,eichen3,witten}. The reductions
derived from the PCM lead to the coset sigma models and in
particular to the symmetric space sigma models \cite{book,mik}.
Therefore from this standpoint PCM is also essential to study a
major branch of the sigma models. On the other hand by taking the
target group manifold as infinite dimensional (large N limit) the
four dimensional self-dual Einstein
\cite{park1,parko,9310003,9402020,9602050,plebanski1,plebanski2}
and also the self-dual Yang-Mills
\cite{parko,park2,park3,manas1,manas2} equations can be formulated
as PCM. On the string theory side the string dynamics on product
manifolds containing Lie group factors lead to PCM with a critical
Wess-Zumino (WZ) term which is needed for conformal anamoly
cancellation in quantizing the theory \cite{witten,witten2}. For
the type IIB Green-Schwarz superstring on $AdS(5)\times S^{5}$ the
field space can be formulated as a coset and the dynamics can be
described via a coset sigma model, a fermionic WZ term and the
$\kappa$ symmetry \cite{gs1,gs2,gs3}. Moreover the bosonic sectors
of the type IIA or IIB superstrings on $AdS_{3}\times S^{3}\times
T^{4}$ can be formulated as worldsheet-PCM \cite{t41,t42} when the
background current is purely R-R otherwise again as a PCMWZ model.
Here we should remark that the non-critical PCMWZ models can
always be derived from the PCM via field transformations based on
the extended solutions of the Lax formalism \cite{sezgin}. The
symmetries of these superstrings as integrable systems have been
studied extensively \cite{bena,07052858}\footnote{We refer the
reader especially to the references in \cite{07052858} for the
integrability aspects of these superstrings.}. Apart from these
superstrings in the low energy limit also nearly all the scalar
sectors of the supergravity theories can be formulated as
symmetric space sigma models \cite{sezgin, julia1,julia2}. Thus as
a result of its appearance in string theory and geometry studying
PCM has become crucial in understanding and possibly refining the
dynamical or the algebraic geometrical features and extending the
structures of the relative models or the theories. We can refer
the reader to a moderate selection from the rich literature
studying different aspects of PCM in
\cite{book,mik,eichen3,eichen5,eichen2,eichen1,sch1,sch2}.

In \cite{prindual} by considering the components of the invariant
right-Noether current of the $D$-dimensional model which has a
global left-right acting $G\times G$ symmetry as independent
fields we have formulated a dual theory. This led us to the
first-order algebraic equations of these current components in
terms of the dual fields which appear in the dual construction as
Lagrange multipliers. In this work we will solve these equations
for the $2D$ case. We will show that the solutions to these
first-order equations exactly determine the form of the
right-current in terms of the dual fields which become scalars in
the $2D$ case. The solution appears to be in the form of a Lax
connection. For this reason we realize that this solution together
with the dual field equation which becomes the zero curvature
condition for this connection define an integrated\footnote{In the
sense that it is explicitly composed of scalar fields rather than
currents.} Lax formalism for the theory. As we obtain the current
components explicitly in terms of the dual fields we will also
present the dual lagrangian whose independent fields are solely
the dual scalars. In Section five we will construct the
first-order PDE system of the theory by equating the explicitly
calculated Noether-currents both in terms of the original
\cite{nej1,nej2,nej3} and the dual scalar fields of the theory for
the case of a lie algebra parametrization of the dynamic
$G$-valued map which constitutes the original lagrangian. Finally
in Section six we will discuss the Frobenious integrability of
these first-order equations.
\section{Currents in terms of Dual Fields}
The first-order field equations of a generic $D$-dimensional PCM
was derived in \cite{prindual} by means of introducing a set of
Lagrange multipliers. They read
\begin{equation}\label{e1}
(-1)^{D}T_{ml}\ast F^{m}=-dA_{l}-C_{ln}^{k}F^{n}\wedge A_{k}.
\end{equation}
Here $\{F^{m}\}$ are the components of the Noether right-current
which is a flat connection of the half $G-$global symmetry of the
model. $\{A_{l}\}$ are the $(D-2)$-forms which are the free
Lagrange multipliers of the theory $D$ being the dimension of the
base manifold which for the superstring case becomes the
worldsheet. When $\{F^{m}\}$ are solved via \eqref{e1} in terms of
$\{A_{l}\}$ the permissable dual fields $\{A_{l}\}$ appearing in
\eqref{e1} are the ones which enable $\{F^{m}\}$ to satisfy the
Bianchi identities
\begin{equation}\label{e3}
dF^{l}=-\frac{1}{2}C_{mn}^{l}F^{m}\wedge F^{n}.
\end{equation}
In other words they must be solved from these equations. As a
consequence of the construction of the dual theory in
\cite{prindual} \eqref{e1} and \eqref{e3} replace the original PCM
field equations. All the Latin indices in the above formulation
run from $1$ to the dimension of the target group $G$.
$C_{ln}^{k}$ are the structure constants of the corresponding Lie
algebra. $T_{ml}$ is the Cartan-Killing metric or more generally
the trace convention of the representation chosen for the Lie
algebra of $G$. In this section we will systematically derive the
solutions of $F$ in terms of $A$ when $D=2$ with signature $s=1$.
In other words we will solve \eqref{e1} for $D=2$. When $D=2$ the
Lagrange multipliers $A$ become scalar fields. First let us
observe that in $D=2$ on a one-form for which $p=1$
\begin{equation}\label{e4}
\ast\cdot\ast=(-1)^{(p(D-p)+s)}=(-1)^{(1(2-1)+1)}=1.
\end{equation}
Before attacking on \eqref{e1} let us first consider a similar
non-matrix equation
\begin{equation}\label{e5}
\ast F =-dA-F\wedge A,
\end{equation}
which does not contain group indices. In solving \eqref{e5} as a
first trial if one assumes
\begin{equation}\label{e6}
F =-\ast dA,
\end{equation}
from \eqref{e5} one obtains
\begin{equation}\label{e7}
-\ast\ast dA =-dA\neq-dA+A\ast dA.
\end{equation}
Then a second trial
\begin{equation}\label{e8}
F =-\ast dA+AdA,
\end{equation}
yields
\begin{equation}\label{e9}
-\ast\ast dA+A\ast dA\neq-dA+A\ast dA-A^{2}dA.
\end{equation}
Adding again the excessive term to the solution anzats
\begin{equation}\label{e10}
F =-\ast dA+AdA-A^{2}\ast dA,
\end{equation}
via \eqref{e5} yields
\begin{equation}\label{e11}
-\ast\ast dA+A\ast dA-\ast\ast A^{2}dA\neq-dA+A\ast
dA-A^{2}dA+A^{3}\ast dA.
\end{equation}
One more trial
\begin{equation}\label{e12}
F =-\ast dA+AdA-A^{2}\ast dA+A^{3}dA,
\end{equation}
gives
\begin{equation}\label{e13}
-\ast\ast dA+A\ast dA-\ast\ast A^{2}dA+A^{3}\ast dA\neq-dA+A\ast
dA-A^{2}dA+A^{3}\ast dA-A^{4}dA,
\end{equation}
which suggests the form of the series that solves \eqref{e5}.
Therefore we will assume that
\begin{equation}\label{e14}
\begin{aligned}
F &=-\ast dA+AdA-A^{2}\ast dA+A^{3}dA-A^{4}\ast dA+A^{5}dA-A^{6}\ast dA+\cdot\cdot\cdot\\
\\
 &=-(1+A^{2}+A^{4}+A^{6}+\cdot\cdot\cdot)\ast
 dA+(A+A^{3}+A^{5}+A^{7}+\cdot\cdot\cdot)dA.
\end{aligned}
\end{equation}
Since for $-1<x<1$
\begin{subequations}\label{e15}
\begin{align}
\frac{1}{1+x}+\frac{1}{1-x}&=2(1+x^{2}+x^{4}+\cdots),\notag\\
\notag\\
\frac{1}{1-x}-\frac{1}{1+x}&=2(x+x^{3}+x^{5}+\cdots),\tag{\ref{e15}}
\end{align}
\end{subequations}
we have
\begin{equation}\label{e16}
F =-\frac{1}{2}(\frac{1}{1+A}+\frac{1}{1-A})\ast
dA+\frac{1}{2}(\frac{1}{1-A}-\frac{1}{1+A})dA,
\end{equation}
which yields
\begin{equation}\label{e17}
F =-\frac{1}{1-A^{2}}\ast dA+\frac{A}{1-A^{2}}dA.
\end{equation}
Although this solution is derived via a series expansion by
inspecting term by term one can show that by simply substituting
\eqref{e17} in \eqref{e5} its functional form satisfies \eqref{e5}
and thus it is a particular solution. However it needs to be
checked whether \eqref{e17} is the most general form of solution
to the algebraic equation \eqref{e5}. Now being equipped with the
form of solutions for the corresponding non-matrix equation we
will focus on \eqref{e1} which for $D=2$ can be written as
\begin{equation}\label{e18}
T\ast F=-dA-CF,
\end{equation}
where we start using the matrix notation with definitions
\begin{equation}\label{e19}
(T)^{m}_{\:\:\:\:\:l}=T_{ml}\quad,\quad
(C)^{l}_{\:\:\:n}=C^{k}_{ln}A_{k}\quad,\quad
(F)^{m}=F^{m}\quad,\quad (A)^{k}=A_{k}.
\end{equation}
Here again $\{A_{k}\}$ are scalar fields. In the following instead
of \eqref{e18} we will prefer to work on
\begin{equation}\label{e20}
\ast F=-T^{-1}dA-T^{-1}CF.
\end{equation}
Now like we have done above the trials of solution
\begin{equation}\label{e21}
\begin{aligned}
F &=-T^{-1}\ast dA,
\\
F &=-T^{-1}\ast dA+T^{-1}CT^{-1}dA,
\\
F &=-T^{-1}\ast dA+T^{-1}CT^{-1}dA-T^{-1}CT^{-1}CT^{-1}\ast dA,
\\
F &=-T^{-1}\ast dA+T^{-1}CT^{-1}dA-T^{-1}CT^{-1}CT^{-1}\ast
dA\\
&\quad+T^{-1}CT^{-1}CT^{-1}CT^{-1}dA,
\end{aligned}
\end{equation}
in \eqref{e20} suggests that the series expansion of the general
solution has the form
\begin{equation}\label{e22}
\begin{aligned}
F &=-T^{-1}(1+(CT^{-1})^{2}+(CT^{-1})^{4}+\cdot\cdot\cdot)\ast
 dA\\
\\
 &\quad+T^{-1}(CT^{-1}+(CT^{-1})^{3}+(CT^{-1})^{5}+\cdot\cdot\cdot)dA.
\end{aligned}
\end{equation}
With the help of the Taylor expansion
\begin{equation}\label{e23}
(1+X)^{-1}=1-X+X^{2}-X^{3}+\cdots,
\end{equation}
for matrices we can express \eqref{e22} as
\begin{equation}\label{e24}
F =-T^{-1}(1-(CT^{-1})^{2})^{-1}\ast
dA+T^{-1}CT^{-1}(1-(CT^{-1})^{2})^{-1}dA.
\end{equation}
Again we have derived this solution as a series expansion however
as before we can once more show that its functional form
\eqref{e24} solves \eqref{e20} exactly \footnote{One may again
check whether \eqref{e24} is the general solution of
\eqref{e20}.}. We can prove this by directly substituting
\eqref{e24} in \eqref{e20}. If one does so the left hand side
(lhs) becomes
\begin{equation}\label{e25}
-T^{-1}(1-(CT^{-1})^{2})^{-1}dA+T^{-1}CT^{-1}(1-(CT^{-1})^{2})^{-1}\ast
dA,
\end{equation}
and the right hand side (rhs) yields
\begin{equation}\label{e26}
-(T^{-1}+T^{-1}CT^{-1}CT^{-1}(1-(CT^{-1})^{2})^{-1}dA+T^{-1}CT^{-1}(1-(CT^{-1})^{2})^{-1}\ast
dA.
\end{equation}
Now by using the binomial inverse theorem
\begin{equation}\label{e27}
(A+B)^{-1}=A^{-1}-A^{-1}B(B+BA^{-1}B)^{-1}BA^{-1},
\end{equation}
one can show that
\begin{equation}\label{e28}
(1-(CT^{-1})^{2})^{-1}=-TC^{-1}TC^{-1}+((CT^{-1})^{2})^{-1}(1-(CT^{-1})^{2})^{-1}.
\end{equation}
Thus when substituted in \eqref{e26} this proves that the rhs
namely \eqref{e26} and the lhs \eqref{e25} become equal.
\section{The Integrated Lax Formulation}
In the previous section we have exactly derived the solutions of
\eqref{e1} which express the components of the Noether current
(which is conserved due to the half of the $G$-global symmetry of
the PCM) in terms of the dual (Lagrange multiplier) fields in
\eqref{e24}. Now referring to \cite{prindual} we should face the
fact that as a coarse of the first-order formulation, in addition
to \eqref{e24} these current components which were considered as
independent fields in \cite{prindual} must also satisfy the second
set of field equations \eqref{e3} of the dualised theory. These
Bianchi identities originate from the zero curvature condition
\begin{equation}\label{ee29}
d\mathcal{G}^{\prime}+ \mathcal{G}^{\prime}\wedge
\mathcal{G}^{\prime}=0,
\end{equation}
of the right-current $\mathcal{G}^{\prime}=F^{m}T_{m}$. Here
$\{T_{m}\}$ are the generators of the Lie algebra of the target
group manifold $G$. By using \eqref{e24} we can express
$\mathcal{G}^{\prime}$ as
\begin{equation}\label{ee30}
\mathcal{G}^{\prime}=\big[T^{-1}CT^{-1}(1-(CT^{-1})^{2})^{-1}dA\big]^{m}T_{m}
-\big[T^{-1}(1-(CT^{-1})^{2})^{-1}\ast dA\big]^{m}T_{m}.
\end{equation}
This relation whose components are the general solutions of
\eqref{e1} together with \eqref{ee29} replace the original
second-order field equations of the PCM. In this section we will
show that \eqref{ee29} and \eqref{ee30} are indeed forming an
integrated Lax formalism for the theory. First let us observe that
\eqref{ee29} has the general solution
\begin{equation}\label{ee31}
\mathcal{G}^{\prime}=g^{-1}dg,
\end{equation}
where $g$ is the Lie group valued map. Independently we also owe
the pure gauge form of \eqref{ee31} to the construction of the
dual theory in \cite{prindual} where we start with directly the
Noether current \eqref{ee31}. At this stage we should remark that
owing to the formulation of \cite{prindual} \eqref{e1} and
\eqref{ee29} are independent of the form or the parametrization of
$g$ thus the results there and here are valid for the entire
solution space. Now let us call
\begin{equation}\label{ee32}
\lambda(A^{n})=CT^{-1},
\end{equation}
which is a matrix function of the dual scalar fields. Substituting
\eqref{ee31} in \eqref{ee30}, and putting it in a more explicit
form, also giving it a new name we get
\begin{equation}\label{ee33}
L=g^{-1}dg=\big[T^{-1}\lambda\frac{1}{1-\lambda^{2}}\big]^{m}_{n}dA^{n}T_{m}
-\big[T^{-1}\frac{1}{1-\lambda^{2}}\big]^{m}_{n}\ast
 dA^{n}T_{m}.
\end{equation}
Here we prefer the notation
\begin{equation}\label{ee34}
\frac{1}{1-\lambda^{2}}\equiv(1-\lambda^{2})^{-1}.
\end{equation}
\eqref{ee33} is certainly a Lax connection whose consistency or
integration conditions namely \eqref{ee29} lead to the field
equations of the theory in the dual language. The reader should
realize that in \eqref{ee33} although $\lambda$ appears at the
place of a spectral parameter which is a complex variable of the
ordinary Lax connection
\begin{equation}\label{ee34}
L=\frac{\lambda}{1-\lambda^{2}}J -\frac{1}{1-\lambda^{2}}\ast
 J,
\end{equation}
where $J$ happens to be the Noether current of the PCM lagrangian.
In \eqref{ee33} it is a matrix functional of the dual scalar
fields. This is a bare consequence of the overall integration of
the system of field equations of the theory by degree one.
Therefore we may simply call \eqref{ee33} an integrated Lax
formalism of the PCM. We can furthermore express \eqref{ee33} as
\begin{equation}\label{ee35}
L=\bigg[\big[T^{-1}\lambda\frac{1}{1-\lambda^{2}}\big]^{m}_{n}\partial_{\alpha}A^{n}T_{m}
-\big[T^{-1}\frac{1}{1-\lambda^{2}}\big]^{m}_{n}\partial^{\beta}A^{n}\varepsilon_{\beta\alpha}T_{m}\bigg]dx^{\alpha},
\end{equation}
where on the $2D$ base manifold $M$ we have introduced
$\varepsilon_{\alpha\beta}=\sqrt{-det(h)}\epsilon_{\alpha\beta}$
with $\epsilon_{01}=1$ and $h$ is the Minkowski signature metric
on $M$. Thus the components of $L$ become
\begin{equation}\label{ee36}
L_{\alpha}=\big[T^{-1}\lambda\frac{1}{1-\lambda^{2}}\big]^{m}_{n}\partial_{\alpha}A^{n}T_{m}
-\big[T^{-1}\frac{1}{1-\lambda^{2}}\big]^{m}_{n}\partial^{\beta}A^{n}\varepsilon_{\beta\alpha}T_{m},
\end{equation}
for which $\alpha=1,2$. Finally we can define the integrated Lax
pair of the theory as
\begin{equation}\label{ee37}
\partial_{\alpha}g=gL_{\alpha},
\end{equation}
whose compatibility conditions
$\partial_{\alpha}\partial_{\beta}g=\partial_{\beta}\partial_{\alpha}g$
which are equivalent to \eqref{ee29} can be expressed in component
form as
\begin{equation}\label{ee37.5}
\partial_{\alpha}L_{\beta}-\partial_{\beta}L_{\alpha}+[L_{\alpha},L_{\beta}]=0.
\end{equation}
As a result of the construction of the dual theory these equations
are equivalent to the original field equations of the PCM.
\section{The Dual Lagrangian}
Now as we have explicitly solved \eqref{e1} for $D=2$ we can again
explicitly construct the lagrangian of the dual theory in terms of
the dual fields $\{A_{k}\}$. The Bianchi-term containing
lagrangian was introduced in \cite{prindual} in terms of
$\{A_{k}\}$ and the flat connection components $\{F^{m}\}$. Since
we have also expressed $\{F^{m}\}$ in terms of the dual scalar
fields the pure form of the lagrangian of the dual scalar field
theory can now be given as
\begin{equation}\label{ee28.6}
{\mathcal{L}}_{Dual}=-\frac{1}{2}\, tr(\ast
\mathcal{G}^{\prime}\wedge
\mathcal{G}^{\prime})+\mathcal{F}^{l}A_{l},
\end{equation}
where $\mathcal{G}^{\prime}$ should be taken as \eqref{ee30} and
the field strength in the Chern-Simon type term is
\begin{equation}\label{ee28.65}
\mathcal{F}=d\mathcal{G}^{\prime}+
\frac{1}{2}[\mathcal{G}^{\prime},\mathcal{G}^{\prime}],
\end{equation}
whose components can be given as
\begin{equation}\label{ee28.7}
\mathcal{F}^{l}=dF^{l}+\frac{1}{2}C_{mn}^{l}F^{m}\wedge F^{n}.
\end{equation}
We can now express \eqref{ee28.6} as
\begin{equation}\label{e29}
\mathcal{L}_{Dual}=-\frac{1}{2}\ast F^{m}\wedge
F^{n}T_{mn}+(dF^{l}+\frac{1}{2}C_{mn}^{l}F^{m}\wedge F^{n})\wedge
A_{l},
\end{equation}
where $F^{l}$ must be read from \eqref{e24}. We may further split
the lagrangian \eqref{e29} into two parts by eliminating the cross
terms coming from the connection components in \eqref{e24}. First
let us introduce the vector notation
\begin{equation}\label{e30}
G=T^{-1}(1-(CT^{-1})^{2})^{-1}dA\equiv MdA,
\end{equation}
so that we can write \eqref{e24} as
\begin{equation}\label{e30.5}
F =-\ast G+T^{-1}CG.
\end{equation}
Now inserting \eqref{e30.5} in the first term in \eqref{e29} after
some algebra we get
\begin{equation}\label{e31}
\begin{aligned}
-\frac{1}{2}\ast F^{m}\wedge F^{n}T_{mn}&=-\frac{1}{2}G^{m}\wedge
\ast G^{n}T_{mn}+\frac{1}{2}C_{mn}G^{m}\wedge
G^{n}+\frac{1}{2}C_{mn}\ast G^{n}\wedge
\ast G^{m}\\
&\quad-\frac{1}{2}(T^{-1}C\ast G)^{m}\wedge (T^{-1}CG)^{n}T_{mn}.
\end{aligned}
\end{equation}
Furthermore the second part of the second term becomes
\begin{equation}\label{e32}
\begin{aligned}
\frac{1}{2}C_{mn}^{l}F^{m}\wedge F^{n}\wedge
A_{l}&=\frac{1}{2}C_{mn}\ast G^{m}\wedge \ast
G^{n}-\frac{1}{2}C_{mn}\ast G^{m}\wedge (T^{-1}C
G)^{n}\\
&\quad-\frac{1}{2}C_{mn}\ast G^{m}\wedge (T^{-1}C
G)^{n}\\
&\quad+\frac{1}{2}C_{mn}(T^{-1}CG)^{m}\wedge (T^{-1}C G)^{n}.
\end{aligned}
\end{equation}
Now since\footnote{We should remark that we freely raise and lower
Latin indices when necessary for a compact notation.}
\begin{equation}\label{e33}
-\frac{1}{2}(T^{-1}C\ast G)^{m}\wedge
(T^{-1}CG)^{n}T_{mn}=\frac{1}{2}C_{mn}\ast G^{m}\wedge (T^{-1}C
G)^{n},
\end{equation}
Combining \eqref{e31} with \eqref{e32} in \eqref{e29} and
simplifying more we can express the dual lagrangian as
\begin{equation}\label{e34}
\begin{aligned}
\mathcal{L}_{Dual}&=-\frac{1}{2}tr(\tilde{G}\wedge \ast
\tilde{G})-\frac{1}{2}tr(\tilde{G}^{\prime}\wedge \ast
\tilde{G}^{\prime})+\frac{1}{2}C_{mn}\tilde{G}^{m}\wedge\tilde{G}^{n}\\
&\quad+\frac{1}{2}C_{mn}\tilde{G}^{\prime
m}\wedge\tilde{G}^{\prime n}-d(\ast
\tilde{G}^{l})A_{l}+d(\tilde{G}^{\prime l}) A_{l}.
\end{aligned}
\end{equation}
Here we have introduced the dual field strengths
\begin{equation}\label{e35}
\tilde{G}=G^{m}T_{m}\quad,\quad
\tilde{G}^{\prime}=(T^{-1}CG)^{m}T_{m},
\end{equation}
with $\tilde{G}^{m}=G^{m}$ and $\tilde{G}^{\prime
m}=(T^{-1}CG)^{m}$.
\section{Field Equations as a First-order PDE System}
In this section we will comment on some methods of solving the PCM
equations from different points of views. Firstly to solve the
dual theory one may simply derive the Euler-Lagrange equations of
the dual lagrangian which we have derived in the previous section.
However this would not bring more ease as like the original PCM
equations one would again obtain second-order field equations.
Another route can be to consider the equivalent system
\eqref{ee29} and \eqref{ee30} or their combined version
\eqref{ee33}. In this respect a direct method of obtaining general
solutions of the dual theory is to substitute \eqref{ee30} in
\eqref{ee29} and seek solutions for $\{A_{n}\}$. By doing so one
gets the equation
\begin{equation}\label{e37}
d(-\ast\tilde{G}+\tilde{G}^{\prime})+(\ast\tilde{G}+\tilde{G}^{\prime})\wedge(\ast\tilde{G}+\tilde{G}^{\prime})=0.
\end{equation}
In fact the system of equations achieved from \eqref{e37} are more
simplified than the field equations which can be obtained from the
dual lagrangian \eqref{e34}. Alternatively one may consider the
first-order Lax pair \eqref{ee37} where as we have mentioned
before the role of the spectral parameter is replaced by a matrix
functional of dual fields. At this point we may discuss that the
inverse scattering method of \cite{mik,shabat} which is based on
the Riemann-Hilbert problem via the spectral parameter can be
adopted by complexifying the dual fields and considering a
spectral functional of the fields of the theory. One may even
generate generalized forms of soliton solutions in this manner.
Moreover other methods can also be derived to solve \eqref{ee37}
as unlike the original Lax pair associated with \eqref{ee34} whose
building block $J$ is yet to be determined the integrated Lax pair
of \eqref{ee37} is exactly resolved in terms of the dual scalar
fields. In one sense these fields can be considered as dynamically
free since they are only subject to the algebraic integrability
conditions of \eqref{ee37}. By using the classical solution
methods one may expect to generate richer classes of solutions
since the degrees of freedom for manipulating the solutions
increase from a single parameter to a functional due to the
integration of the system by degree one. Another route to seek
solutions is to construct the first-order system of partial
differential equations of the theory for a particular
parametrization of the Lie group valued map $g$. If we consider
the parametrization
\begin{equation}\label{e37.5}
g=e^{\varphi^{m}(x)T_{m}},
\end{equation}
of $g$ with scalars $\varphi^{i}$ then we have
\cite{nej1,nej2,nej3}
\begin{equation}\label{e38}
F^{m}=W^{m}_{\:\:\: n}(\varphi^{l})d\varphi^{n},
\end{equation}
where the dim$G\times$dim$G$ matrix $W$ is
\begin{equation}\label{e38.5}
W=(I-e^{-M})M^{-1}.
\end{equation}
Here we define the matrix $M$ as
\begin{equation}\label{e38.6}
M_{\:\:\:m}^{n}=C_{lm}^{n}\varphi^{l}.
\end{equation}
By definition the rhs of \eqref{e38} automatically satisfies the
zero curvature conditions \eqref{e3}. Substituting \eqref{e38} in
\eqref{e30.5} one obtains
\begin{equation}\label{e39}
W^{m}_{n}(\varphi^{l})d\varphi^{n}=-\ast G^{m}+(T^{-1}CG)^{m}.
\end{equation}
Now one is entirely free to start with any choice of $\{A_{n}\}$,
the unique constraint on them is the existence of a transformation
from the fields $\{\varphi^{m}\}$ to the dual ones $\{A_{n}\}$ via
\eqref{e39}. Thus if one can find explicit or implicit relations
between $\{\varphi^{m}\}$ and $\{A_{n}\}$ which will satisfy
\eqref{e39} then one automatically by-passes \eqref{e3} since the
lhs of \eqref{e39} trivially satisfies \eqref{e3} so does the rhs
due to the constructed transformation. Such a field transformation
method which is seeking algebraic relations between matrix
exponential and matrix polynomial forms may partially replace the
differential equation solving methodology of the original $D=2$
PCM. Furthermore starting from \eqref{e39} we can also construct
the first-order partial differential equation system of the
theory. First let us assume a local coordinate bases
$\{dx^{1},dx^{2}\}$ on the two-dimensional base manifold $M$. We
have already introduced the Minkowski signatured pseudo-Riemannian
metric $h$ on $M$ it can be expressed as
\begin{equation}\label{e40}
h=h_{\alpha\beta}dx^{\alpha}\otimes dx^{\beta},
\end{equation}
where $\alpha,\beta=0,1$. We have \cite{thring}
\begin{equation}\label{e40.1}
\ast dx^{0}=h^{\prime}(h^{00}dx^{1}+h^{01}dx^{0})\quad ,\quad \ast
dx^{1}=h^{\prime}(h^{10}dx^{1}+h^{11}dx^{0}),
\end{equation}
where $h^{\prime}=\sqrt{|Det(h)|}$. Now \eqref{e39} becomes
\begin{equation}\label{e40.2}
\begin{aligned}
&\big[W^{m}_{n}\partial_{0}\varphi^{n}+h^{\prime}M^{m}_{n}\partial_{0}A^{n}h^{01}+h^{\prime}M^{m}_{n}\partial_{1}A^{n}h^{11}-(T^{-1}C)^{m}_{n}M^{n}_{l}
\partial_{0}A^{l}\big]dx^{0}+\\
&+\big[W^{m}_{n}\partial_{1}\varphi^{n}+h^{\prime}M^{m}_{n}\partial_{0}A^{n}h^{00}+h^{\prime}M^{m}_{n}\partial_{1}A^{n}h^{10}-(T^{-1}C)^{m}_{n}M^{n}_{l}
\partial_{1}A^{l}\big]dx^{1}=0.
\end{aligned}
\end{equation}
Here we have used the $M$-matrix notation introduced in
\eqref{e30}. \eqref{e40.2} is a first-order partial differential
equation system for the fields
$\{\varphi^{m}(x^{\alpha}),A^{n}(x^{\alpha})\}$. We should state
that the dual fields enter into this system as free fields without
any additional constraints the integrability conditions of
\eqref{e40.2} will bring algebraic constraints on them which will
also shape the solution space of the dual theory.
\section{Frobenius Integrability}
In this section we will inspect the Frobenius integrability
conditions of \eqref{e39} and we will present the special set of
solutions which follow these conditions. Now starting from
\eqref{e39} let us construct
\begin{equation}\label{e40.4}
W^{m}_{n}(\varphi^{l})d\varphi^{n}T_{m}=(-\ast
G^{m}+(T^{-1}CG)^{m})T_{m}=\mathcal{G}^{\prime}(A_{n}).
\end{equation}
If now we assume that $\mathcal{G}^{\prime}(A_{n})$ is closed
\begin{equation}\label{e40.5}
d\mathcal{G}^{\prime}=0,
\end{equation}
then locally it is an exact form namely there exists a Lie
algebra-valued function $\omega(A_{n})$ such that
\begin{equation}\label{e40.55}
\mathcal{G}^{\prime}(A_{n})=d\omega(A_{n}).
\end{equation}
For $\mathcal{G}^{\prime}(A_{n})$ to be exact we must formally
have
\begin{equation}\label{e40.6}
\begin{aligned}
&\partial_{\mu}(\big[T^{-1}\lambda\frac{1}{1-\lambda^{2}}\big]^{m}_{n}\partial_{\alpha}A^{n}
-\big[T^{-1}\frac{1}{1-\lambda^{2}}\big]^{m}_{n}\partial^{\beta}A^{n}\varepsilon_{\beta\alpha})\\
\\
&=\partial_{\alpha}(\big[T^{-1}\lambda\frac{1}{1-\lambda^{2}}\big]^{m}_{n}\partial_{\mu}A^{n}
-\big[T^{-1}\frac{1}{1-\lambda^{2}}\big]^{m}_{n}\partial^{\beta}A^{n}\varepsilon_{\beta\mu}).
\end{aligned}
\end{equation}
However we should also state that instead of solving these
equations by inspection one may seek special forms of closed
1-forms in \eqref{e40.4} by assuming special combinations of
$\{A_{n}\}$. Apart from \eqref{e40.6} if furthermore
$\mathcal{G}^{\prime}$ satisfies
\begin{equation}\label{e42}
[\omega,\mathcal{G}^{\prime}]=0,
\end{equation}
then by applying an exterior derivative on \eqref{e42} one can
show that
\begin{equation}\label{e43}
\mathcal{G}^{\prime}\wedge\mathcal{G}^{\prime}=0.
\end{equation}
In this case via \eqref{e40.5} and \eqref{e43} we have
$d\mathcal{G}^{\prime}+\mathcal{G}^{\prime}\wedge\mathcal{G}^{\prime}=0$
which being the zero curvature condition \eqref{ee29} is the
second set of equations of the dual form of the PCM. Thus the set
of fields $\{A_{n}\}$ which would satisfy \eqref{e40.6} and
\eqref{e42} are the solutions of the dual theory. Furthermore
since \cite{prindual}
\begin{equation}\label{e44}
\begin{aligned}
W^{m}_{n}d\varphi^{n}T_{m}&=e^{-\varphi^{m}T_{m}}de^{\varphi^{n}T_{
n}}\\
&=d(\varphi^{m}T_{m})-\frac{1}{2}[\varphi^{
m}T_{m},d(\varphi^{n}T_{n})]+\frac{1}{6}[\varphi^{m}T_{m},[\varphi^{n}T_{n},d(\varphi^{l}T_{l})]]-\cdots,
\end{aligned}
\end{equation}
if one finds fields $\{A_{n}\}$ which satisfy equations
\eqref{e40.6} and \eqref{e42} then choosing
\begin{equation}\label{e45}
\varphi^{m}=\omega^{m},
\end{equation}
with $\omega=\omega^{m}T_{m}$ gives
\begin{equation}\label{e46}
W^{m}_{n}d\varphi^{n}=d\omega^{m}=\mathcal{G}^{\prime m},
\end{equation}
via \eqref{e44} where due to the choice of \eqref{e45} \eqref{e42}
can be used to eliminate all the terms but the first. Under these
conditions as \eqref{e46} is the same equation with \eqref{e40.4}
the first set of equations of the theory are also satisfied. Thus
\eqref{e45} become the solutions of the original $D=2$ PCM. In
this respect equations \eqref{e40.6} and \eqref{e42} give the dual
solutions whereas by calculating $\mathcal{G}^{\prime}$ and then
$\omega^{m}$ via \eqref{e40.55} in terms of these dual solutions
\eqref{e45} provide the original solutions of the Lie algebra
parametrization of the map $g$.
\section{Conclusion}
In this work we have solved the algebraic first-order equations of
the dual formulation of the PCM which were derived in
\cite{prindual}. Therefore by expressing the right-Noether current
in terms of the dual fields we have obtained an integrated Lax
formulation of the theory. We call it integrated since the
formulation presented in \cite{prindual} and the solutions derived
here enable us to reduce the degree of the system of equations by
one. This reflects itself in the Lax connection and the associated
Lax pair. The ordinary Lax pair contains a complex spectral
parameter however although the usual form of the Lax connection
again appears here this time the constant parameter is substituted
with a matrix functional of the fields of the dual theory.
Furthermore we have also taken the advantage of reaching the above
mentioned solutions in constructing the dual lagrangian which is
the most general dual scalar field theory equivalent of the PCM.
Following a concise discussion on the system of equations of the
dual theory, by the means of the matrix exponential-polynomial
transformation of the current components to the scalar field
strengths for the exponential parametrization of the group map we
have mentioned various solution techniques, and also obtained the
first-order PDE system of the $D=2$ PCM. In connection with this
we have discussed the Frobenious integrability of this system.

The $D=2$ PCM has a deeper and a more extensive symmetry scheme
than the global $G\times G$ left-right symmetry. Firstly as an
integrable system it has infinitely many local left-right
conserved charges \cite{chered1,chered2,chered3,
eichen6,loc1,loc2,loc3,kay1,kay2,kay3} which form $W$-algebras
\cite{kay1,kay2}. Moreover apart from these usual symmetries of
the integrable systems there are non-local conserved charges which
extend the $G\times G$ global symmetry algebras to the infinite
dimensional Kac-Moody algebras which lead to the Yangians at the
quantum level as well as the Virasoro algebras which on the total
are called the hidden symmetries. The Lax formalism is at the
heart of studying these symmetries. Starting from the Lax pair
which serves as the generating functional for the monodromy
matrices one studies the Poisson brackets of these matrices to
obtain the Yang-Baxter type infinite algebras for the ultra-local
cases and the more general ones for the non-ultra-local cases
which are both related to the Kac-Moody algebras
\cite{eichen3,eichen5,pohlmeyer,mon1,mon2,mon3,mon4,mon5,mon6,mackay92,eichen4,maillet1,forger1,maillet2,maillet3}.
In either case in constructing the infinite dimensional symmetry
algebras the explicit form of the Lax pair and the inverse
scattering method of \cite{mik,shabat} play the central role. In
the ultra-local models, field independent $r$-matrices appear in
the commutation relations of the spatial parts of the Lax pair
whereas in the more generalized non-ultra-local models the field
dependent $s$-matrices accompany the $r$-matrices. In addition the
monodromy matrices are also defined as the ordered exponentials of
the integrals of these Lax pair components satisfying similar
Poisson brackets. Moreover an infinite number of conserved charges
can be derived from the spectral parameter mode expansions of the
monodromy matrices with the regularization provided by the
infinite volume limit of the inverse scattering method.
Equivalently in another direction the same Kac-Moody symmetries
can be realized as infinitesimal transformation algebras (loop
algebras) acting on the solution space of the PCM
\cite{onsol1,onsol2,onsol3,onsol4,onsol5,onsol6,onsol7,onsol8,onsol9,onsol10}.
In either of these approaches the hierarchy of continuity
equations of the infinitely many global symmetries or the
infinitesimal transformations are direct consequences of the Lax
equation. On the other hand the energy-momentum tensor as a result
of the backstage canonical structure also gives rise to Virasoro
symmetries \cite{viro1,viro2,viro3,viro4,viro5} for the PCM. In
recent years in connection with the research programme on the
integrability of the related string models the hidden symmetries
of the PCM have been revisited
\cite{sezgin,07052858,sch1,sch2,kay1,kay2,kay3,mackay92,devchand-schiff1,devchand-schiff2,kluson}\footnote{The
reader may find a complete historical survey of the hidden
symmetries of the PCM in \cite{sch1}.}. The supersymmetric
extensions of the PCMWZ are also studied
\cite{kay1,kay3,susywz1,susywz2,susywz3,susywz4,susywz5,susywz6}\footnote{
Which for example appear in the strings on group manifolds with
world-sheet supersymmetry \cite{susy1}.}. We may easily state that
for all these above mentioned hidden symmetries the underlying
framework is a natural consequence of the two distinct Hamiltonian
formulations and the associated canonical structures provided by
the Lax pair. Moreover the algebra structures as well as the
explicit representations of the building blocks of these
symmetries depend on the explicit form of the Lax connection
components.

Therefore we may conclude that the Lax formalism is at the center
of exploring the hidden symmetries of the PCM. On the other hand
in this work we have derived a Lax connection and a Lax pair for a
general base manifold and its Minkowski signature
pseudo-Riemannian structure barely in terms of the fields of the
dual theory by performing an overall integration of the field
equations. Although the form of this integrated Lax formulation
resembles the ordinary one (on which the entire above-mentioned
analysis is based on) in our formulation the role of the spectral
parameter is replaced by a matrix functional of the dual fields.
Therefore following a complexification the degrees of freedom of
the spectral analysis can be made a continuum. By means of these
spectral matrix functionals the dynamics of the theory may
directly enter into the canonical structure whose implications can
be read from the field-dependent $s$-matrices of the more
generalized non-ultra-local cases. At this point we should state
that an integrated Lax pair may lead to a more refined and
generalized hidden symmetry formulations as it is one degree of
derivation close to the general solutions of the theory. Besides
the explicit form of the connection components will enable one to
compute explicitly the elements and the representations of the
infinite-dimensional symmetry algebras. In this respect one may
hope to find extensions of the already existing structures in the
literature. Further mode expansions in the fields of the spectral
matrix functional will certainly lead to richer structures. Apart
from this new form of integrated Lax connection our results also
contribute simplifications to solution generation. The approach of
Section five reduces the problem of finding general solutions of
the theory into finding field transformations between the original
and the dual fields. This requires to find implicit or explicit
forms of transformations relating the matrix exponential and
polynomial dressings of the differentials on either side. Apart
from being a tool to search for general solutions the first-order
form of the field equations may also provide a framework for the
study of the transformations of the solutions such as the Backlund
ones and their possible generalizations. One may also hope to
build new solution generating techniques other than the inverse
scattering method built on the new Lax formalism we have derived
as we have discovered a new notion for the spectral degrees of
freedom. Before ending we should remark the resemblance of the
hierarchy of mode expansions \cite{sezgin} and the sequence of the
solution ansatz namely the Taylor expansion terms of the
coefficients of our integrated Lax connection. We also believe
that the integrated dual Lax formalism constructed here should be
covering the special cases of the T-dual string Lax formulations
studied in \cite{son0,son1,son2}. One may also use the formalism
introduced here to extend and generalize the results of
\cite{devchand-schiff1} which studies the solutions of the $U(N)$
PCM. Finally further research may be performed to derive similar
derivations of Section two and three for the supersymmetric
extensions of PCM as well as the PCMWZ models.

\end{document}